\documentclass[graybox]{svmult}

\usepackage{type1cm}        % activate if the above 3 fonts are
\usepackage{makeidx}         % allows index generation
\usepackage{graphicx}        % standard LaTeX graphics tool
\usepackage{multicol}        % used for the two-column index
\usepackage[bottom]{footmisc}% places footnotes at page bottom
\usepackage{pdfpages}

\usepackage{newtxtext}       % 
\usepackage{newtxmath}       % selects Times Roman as basic font
\usepackage{float}
\usepackage{amsfonts}
\usepackage[firstabbrev]{authorindex}
\usepackage{url}
\usepackage{natbib}
\usepackage{verbatim}
\usepackage{algorithm}
\usepackage{algpseudocode}
\usepackage{enumitem}
\usepackage{amsmath}
\usepackage{multirow}
\DeclareMathOperator*{\argmax}{argmax}
\DeclareMathOperator*{\argmin}{argmin}

\makeindex             % used for the subject index
\begin{document}

\title*{Large-Scale Shrinkage Estimation under Markovian Dependence}
\author{Bowen Gang,  Gourab Mukherjee and Wenguang Sun}
\institute{Bowen Gang \at University of Southern California, 3620 S. Vermont ave, 
Los Angeles, CA 90089, USA \email{bgang@usc.edu}
\and Gourab Mukherjee \at University of Southern California,  3670 Trousdale Pkwy, Los Angeles, CA 90089, \email{gourab@usc.edu}
\and Wenguang Sun \at University of Southern California,  3670 Trousdale Pkwy, Los Angeles, CA 90089, \email{wenguangs@marshall.usc.edu}
}
\maketitle

\abstract*{We consider the problem of simultaneous estimation of a sequence of dependent parameters that are generated from a hidden Markov model. Based on observing a noise contaminated vector of observations from such a sequence model, we consider simultaneous estimation of all the parameters irrespective of their hidden states under square error loss.  We study the roles of statistical shrinkage for improved estimation of these dependent parameters. Being completely agnostic on the distributional properties of the unknown underlying Hidden Markov model, we develop a novel non-parametric shrinkage algorithm. Our proposed method elegantly combines \textit{Tweedie}-based non-parametric shrinkage ideas with efficient estimation of the hidden states under Markovian dependence. Based on extensive numerical experiments, we establish superior performance our our proposed algorithm compared to non-shrinkage based state-of-the-art parametric as well as non-parametric algorithms used in hidden Markov models. We provide decision theoretic properties of our methodology and exhibit its enhanced efficacy over popular shrinkage methods built under independence. We demonstrate the application of our methodology on real-world datasets for analyzing of temporally dependent social and economic indicators such as search trends and unemployment rates as well as estimating spatially dependent Copy Number Variations.}

\section{Introduction}
\label{sec:Intro}

%To effectively solve large-scale multiple estimation problem, it is necessary to incorporate  notions of shrinkage. The concept of shrinkage is important because it provides an elegant framework for combining information from related populations and often leads to substantial improvements in the performances of estimators in simultaneous inference. 
Shrinkage is a useful notion that provides an elegant and powerful framework for compound estimation problems. The topic has been extensively studied since the celebrated work of \citet{james1961estimation}. A plethora of influential results obtained by various researchers show that shrinkage often leads to substantial improvements in the performances of conventional estimators. The optimal directions and magnitudes of shrinkage estimators in parametric models have been extensively studied in the literature, see \cite{wells-book,Johnstone-book,ahmed2012empirical,carlin2010bayes} for reviews of related topics. In real-world applications, parametric families of estimators often have limited usage. Nonparametric shrinkage methods, exemplified by Tweedie's formula (\citealp{efron2011tweedie}), have received renewed attention in large-scale inference problems where thousands of parameters are estimated simultaneously; see \cite{brown2009, jiang2009,koenker2014convex,saha2017nonparametric,dicker2016high} for some important recent developments. 

One essential limitation of existing non-parametric shrinkage methods is that most assume that observations are independently sampled from an underlying distribution. However, observations arising from large-scale estimation problems are often dependent. Ignoring the dependence structure may result in significant loss of efficiency and invalid inference. This article aims to develop non-parametric shrinkage estimators under the widely used hidden Markov model and show that the efficiency of existing nonparametric methods can be greatly improved by incorporating the Markovian dependence structure. 

%We focus on non-parametric estimation of a sequence of parameters that share a Markovian dependence structure. 
Consider the problem of simultaneous estimation of a sequence of dependent parameters that are generated from a Hidden Markov Model (HMM) \citep{rabiner1989tutorial,churchill1992hidden}. Markovian dependence provides a powerful tool for modeling data arising from a wide range of modern scientific applications where parameters of interest are spatially or temporally correlated. This article focuses on an important class of HMMs that have two underlying states: one being the default (null) state where the process is \textit{in-control}; the other being the abnormal (non-null) state where the process is \textit{out-of-control}. The observed data are independent conditional on the unknown states. The two-state HMM is a popular model that can be used to describe many data sets collected from various applications. For example, in estimating the Copy Number Variations (CNV) across a genome \citep{efron2011false,jiang2015codex}, there are regions with high and low variations that can be well described by the non-null and null states, respectively. In event analysis application, such as estimating hourly social media trends of a marketing campaign there are dormant (null) and active (non-null) states which can be adaptively incorporated into an HMM for building a better tracker.  

%Based on observing a noise contaminated vector of observations from such a sequence model, we consider simultaneous estimation of all the parameters irrespective of their hidden states under square error loss. We study the roles of statistical shrinkage for improved estimation of these dependent parameters. Being completely agnostic on the distributional properties of the unknown underlying HMM, 
To the best of our knowledge, the important problem of nonparametric shrinkage estimation under  dependence has not been studied in the literature. The key idea in our methodology is to combine the ideas in the elegant \textit{Tweedie's formula} \citep{brown2009,jiang2009, efron2011tweedie} and the forward-backward algorithms in HMMs to infer the underlying states, which is further utilized to facilitate fast and robust estimation of the unknown effect sizes (mean parameters) under Markovian dependence. We establish  decision-theoretic properties of the proposed estimator and exhibit its enhanced efficacy over popular shrinkage methods developed under the independence assumption. In contrast with existing algorithms in HMMs that proceed with pre-specified families of  parametric densities (e.g. Gaussian mixtures), our method is nonparametric and capable of handling a wider class of distributions. Through extensive numerical experiments, we demonstrate the superior performance of our proposed algorithm over existing state-of-the-art methods.  

The chapter is organized as follows: in Section 2 we formulate the problem, develop an oracle estimator and compare this estimator with classical shrinkage estimators that do not incorporate HMM structure. In Section 3, we propose a data-driven procedure that mimics the oracle estimator and discuss how to overcome the new difficulties and challenges in the non-parametric approach. In Section 4, we establish theoretical properties of our proposed estimator and show that the nonparametric approach offers substantial efficiency gain over HMM algorithms employing Gaussian mixture models. In Section 5, we present numerical experiments to compare our estimator with existing methods. Section 6 applies the proposed method to three real data examples. Section 7 concludes the article with a discussion on interesting extensions.

\section{Shrinkage estimation in a hidden Markov model}

Consider a two-state hidden Markov model. Suppose we are interested in estimating the mean vector $\pmb\mu=(\mu_1, \cdots, \mu_n)$ based on observed output $\pmb X=(X_1, \cdots, X_n)$. 
In HMMs, the observed data $\pmb X$ can be viewed as a contaminated version of the effect sizes $\pmb\mu$, where the contaminations are white noises following a zero-mean normal distribution with known variance $\sigma^2$. The HMM further assumes that $\mu_i$s are independently distributed conditional on the unknown states $\pmb\theta=(\theta_1, \cdots, \theta_n)\in \{0, 1\}^n$, where $\theta_i$ are Bernoulli variables forming a Markov chain. The latent state $[\theta=0]$ usually represents the null or the in-control state and $[\theta=1]$ corresponds to the non-null or the out-of-control state. For example, in CNV studies,  $X_i $ is the observed number of repeats in the genome, $ [\theta_i=0]$ represents healthy part of genome, and $[\theta_{i}=1]$ represents the part of the genome that is susceptible to perturbation in diseased patients.

\subsection{Model and notations}\label{sec.model}

Formally, the data generation process can be described by a hierarchical model:
\begin{equation}\label{eq.model}
X_i=\mu_i+\epsilon_i,  \text{ for } i=1,\ldots,n  \text{ and } \epsilon_i\stackrel{iid}{\large \sim} N(0, \sigma^{2}), 
\end{equation}
where given $\theta_i$,  $\mu_i$ are conditionally independent following unknown prior distributions  $g_0$ and $g_1$:
$$
\mu_i|\theta_i=0 \overset{iid}{\sim} g_0 \text{ and } \mu_i|\theta_i=1 \overset{iid}{\sim} g_1.
$$
In practice, we often assume that the null distribution $g_0$ is either known or can be modeled by a pre-specified parametric family\footnote{We shall assume that $ f_{0} $ is in certain parametric form. However, the parameters of $f_0$ are not necessarily to be known. This is a reasonable assumption, as in real world applications, researchers often have reasonably good ideas about the behavior of an observation from the in-control state.}. By contrast, we do not specify any parametric forms on $g_1$ because we usually do not have sufficient knowledge on the non-null process $g_1$, which may be difficult to describe using any pre-specified parametric families.  The setting is reasonable for a wide range of application scenarios.

\begin{figure}[H]\label{fig-1}
	\centering
	\includegraphics[scale=0.25]{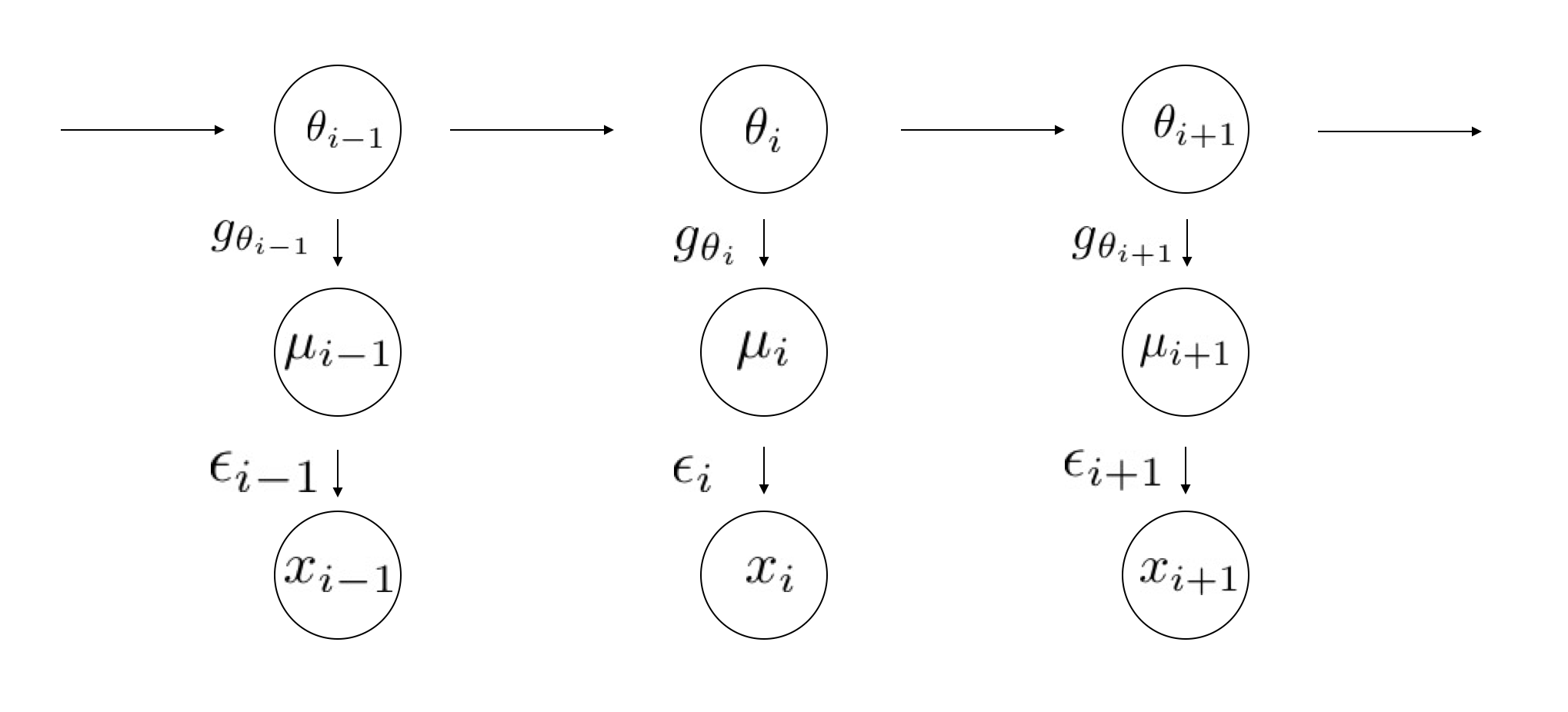}
		\caption{Schematic diagram of the data generation process. We observe $x_i$s only and would like to estimate the $\mu_i$s which are related through the Markov chain $\{\theta_1, \theta_2, \cdots, \theta_n\}$.}
\end{figure}

We assume that $\theta_i$s follow a spatially homogeneous Markov process with transition probabilities 
$$
a_{jk}=P(\theta_i=k|\theta_{i-1}=j), j, k \in \{0,1\}.
$$ 
The above probabilities are unknown and obey standard stochastic constraints
$0<a_{jk}<1, a_{j0}+a_{j1}=1.$
Figure~\ref{fig-1} presents a schematic diagram of the model. 
We aim to find a decision rule $ \pmb{\hat{\mu}}$ for estimating the unknown mean vector $\pmb{\mu}$. The goal is to minimize the Bayes risk $R=E\{l(\hat{\pmb\mu}, \pmb\mu)\}$, where $l(\hat{\pmb\mu}, \pmb\mu)$ is the mean squared error (MSE, or $l_2$ loss): 
$$
l(\hat{\pmb\mu}, \pmb\mu)=n^{-1}\sum_{i=1}^n \mathbb{E}({\mu}_i-{\hat{\mu}_i})^2.
$$

Let $\Lambda=(\mathcal{A},(\psi_{0},\psi_{1}), f_{0},f_{1})$ denote the HMM parameters, where $ \mathcal{A} $ is the transition matrix, $ (\psi_{0},\psi_{1}) $ is the initial distribution, and  
\begin{equation}
f_k(x):=f(x|\theta=k)=\int \phi_\sigma(x-\mu)g_k(\mu)d\mu. \quad k=0, 1.\nonumber
\end{equation} 
We use $\phi_{\sigma}$ to denote the normal density with mean 0 and variance $\sigma^2$ with the suffix dropped for standard normal density.

\subsection{Oracle estimator under independence}

Ignoring the dependence structure, the optimal solution that minimizes the Bayes risk is given by \emph{Tweedie's formula}:
\begin{equation}\label{eq.TND}
\mathbb{E}(\mu_{i}|x_{i})=x_{i}+\sigma^{2}\dfrac{f'(x_{i})}{f(x_{i})}, \text{ where } f'(x)=\frac{d}{dx}f(x) 
\end{equation}
and $f$ is the marginal distribution of all the observed $X$. The formula first appeared in \cite{robbins1955empirical}, who attributed the idea to Maurice Kenneth Tweedie. {Tweedie's formula} can be implemented using empirical Bayes methods for constructing a class of non-parametric estimators \citep{brown2009, efron2011tweedie}. The crucial observation is that it works directly with the marginal distribution, which is in particular attractive in large-scale estimation problems since $f$ and $f'$ can be well estimated using standard density estimation techniques \citep{silverman2018density}.

%For the current problem there are actually two different populations of $\mu$'s with distribution $ g_{0}$ and $ g_{1}$.  Using information from only one population will be better for estimating $ \mu_{i}$. To estimate the underlying hidden state, 

\subsection{Oracle Estimator under HMM dependence}

Our idea is to extend Tweedie's formula to correlated observations under the HMM dependence. This can effectively increase the statistical efficiency by borrowing strength from adjacent observations. It has been shown in the multiple testing literature that exploiting the dependence structure can greatly improve the power of existing false discovery rate \citep{BenHoc95} methods \citep{sun2009large, wei2009multiple,sun2015false}. We further show in this article that dependence structure is also highly informative and can be exploited to improve the accuracy of shrinkage estimators.

Our methodological development is divided into two steps. We first assume that an oracle knows the hidden states and study the oracle rule. Then we discuss the case when the states are unknown and propose data-driven methods to emulate the oracle rule. If the states $\theta_i$ are known, then it is natural to apply Tweedie's formula to two separate states:
\begin{align}\label{eq.OR}
\tilde{\mu_{i}}^{\sf OR}(\pmb{x}) = \sum_{k=0}^1\left\{x_i+\sigma^2\frac{f_k^{'}(x_i)}{f_k(x_i)}\right\} I\{\theta_i=k\}. 
\end{align}
The oracle estimator \eqref{eq.OR} $\tilde{\mu_{i}}^{\sf OR}(\pmb{x})$ provides a benchmark in shrinkage estimation, i.e. the theoretically achievable lower limit of the estimation risk.

Next we consider a ``weaker'' oracle estimator that only knows $\Lambda$, the collection of hyper-parameters in the HMM. Then the optimal solution is given by the next lemma.

\begin{lemma}  
Consider Model  \eqref{eq.model} and assume that HMM parameters are known. Then the estimator that minimizes the MSE is given by
\begin{eqnarray}\label{Bayes-estimator}
\hat {\mu_{i}}^{\sf Bayes}(\pmb{x};\Lambda) & = & \sum_{k=0}^1\mathbb{E}(\mu_i|\theta_i=k, x_i)\mathbb{P}(\theta_i=k|\pmb x)\nonumber
 	\\ & =  & \sum_{k=0}^1\left\{x_i+\sigma^2\frac{f_k^{'}(x_i)}{f_k(x_i)}\right\} \mathbb{P}(\theta_i=k|\pmb x). 
 \end{eqnarray}
 \end{lemma}

The proof of lemma is straightforward and hence omitted. The formula \eqref{Bayes-estimator} can be viewed as an extension of Tweedie's formula under HMM dependence. The next section considers the case where HMM parameters are unknown. We shall develop data-driven estimators and computational algorithms to emulate the Bayes estimator \eqref{Bayes-estimator}.

\section{Data-Driven Estimator and Computational Algorithms}

We discuss how to estimate $f_{0}, \ f_{1}  $ and $ \mathbb{P}(\theta_i=k|\pmb x)$ to implement the Bayes estimator. The proposed estimator is a non-parametric \textit{Tweedie}-based shrinkage estimator under \textit{dependence} (TD). In light of \eqref{Bayes-estimator}, we consider a class of estimators in the form: 
\begin{equation}\label{eq.TD}
\hat{\mu}^{\sf TD}_i=\hat{p}_i\left\{x_i+ \sigma^2\dfrac{\hat{f}_1^{'}(x_i)}{\hat{f}_1(x_i)}\right\}+(1-\hat{p}_i)\left\{x_i+\sigma^2\dfrac{\hat{f}_0^{'}(x_i)}{\hat{f}_0(x_i)}\right\},
\end{equation}
where $\hat p_i$ are estimates of the true conditional probability $P(\theta_i=1|X)$, and $\hat f_0$ and $\hat f_1$ are estimates of $f_0$ and $f_1$. Next we describe an algorithm for constructing estimates of the unknown quantities in \eqref{eq.TD}. 

\subsection{The modified Baum-Welch algorithm} 

The most well-known method for constructing estimators of the form \eqref{eq.TD}, under the conventional setting with pre-specified parametric families, is the Baum-Welch (BW) algorithm \citep{rabiner1989tutorial,yang2015statistical,baum1970maximization}. We emphasize that the BW algorithm must proceed with user specified parametric forms for $f_k$'s. However, while $f_0$, the in-control $(\theta=0)$ distribution, can be well modeled by parametric densities, the out-of-control observations may not be easily approximated by parametric densities. For example, the Gaussian mixture model with a fixed number of components is not suitable for approximating very heavy tailed $f_1$. To overcome the issue, we describe an HMM-based Tweedie (HMMT) estimator that employs parametric $f_0$ and non-parametric $f_1$. The essential component in our estimator is a \emph{generalized BW algorithm} that updates the estimates of $f_0$ and $f_1$ iteratively based on posterior probability estimates of the latent states.

To avoid the identifiability issues, assume $f_0$ is unimodal and $\pi_0 > 0.5$. These assumptions are usually reasonable in applications. In our HMMT estimator, $f_0$ is assumed to be a Gaussian density $\phi[\nu, \tau]$ with possibly unknown location $\nu$ and scale $\tau$. The Gaussian model can be generalized to Gaussian mixtures as well as other parametric families without essential difficulty. In contrast with existing HMM algorithms that utilize parametric assumptions on $f_1$, we consider a \emph{nonparametric} approach to estimating $f_1$. The proposed methodology  employs a class of weighted kernel density estimators
\begin{equation}\label{eq.f1.hat}
\hat{f_{1}}[w,h](x)= \sum_{i=1}^{n}\dfrac{w_{i}}{h}K\bigg(\dfrac{x_{i}-x}{h}\bigg),
\end{equation} 
where $\sum_{i=1}^n w_i=1$, $h$ is the bandwidth and $K(\cdot)$ is a standard Gaussian kernel function. Other choices of kernel such as the Cauchy kernel can also be used particularly in applications that are sensitive to the tails of the density.

With an initial choice of $\mathcal{A}^{(0)}$, $\hat f_0^{(0)}$ and $\hat f_1^{(0)}$, the forward-backward propagation steps of the Baum-Welch algorithm provides updated estimates of the posterior probabilities $\hat p_i^{(0)}$. These estimates can be utilized to update $\hat f_1$ using new weights $w_i^{(0)}=\hat p_i^{(0)}/\sum_{i=1}^n \hat p_i^{(0)}$ and $\hat f_0$: 
\begin{equation*}
\hat{f_{1}}^{(1)}= \hat f_1[\hat w^{(0)},h],  \text{ and } 
(\nu^{(1)}, \tau^{(1)})=\argmax_{\nu \in \mathbb{R}, \tau \geq 0} \sum_{i=1}^n \big(1-\hat p_i^{(0)}\big) \log \phi_{\tau}(X_i-\nu) 
\end{equation*} 
The process is repeated until convergence. 

\subsection{Choice of $h$} 

Choosing an appropriate bandwidth $h$ is crucial for constructing an efficient estimator. If $h$ is too small, then we will end up with estimating $ \mu_{i} $ by the naive estimator $ x_{i}$. If $h$ is too large, then it will be difficult to identify $\hat f_{1}$, which leads to a severely biased and inaccurate estimator. This section describe a cross validation method for choosing the tuning parameter $h$.

The first step is to split the observed sample $\pmb{X}$ into $(\pmb{U},\pmb{V})$ by artifically adding independent Gaussian noise: $\pmb U=\pmb X+\alpha\pmb Z,\ \pmb V=\pmb X-(1/\alpha)\pmb Z$, where $ \pmb Z\stackrel{d}{=} N(0,I_n)$. Note that $\pmb{U}, \pmb{V}$ are normally distributed random effects with same mean $\pmb{\theta}$ but different variances $\sigma_u^2=(1+\alpha^2)\sigma^2$ and $\sigma_v^2=(1+\alpha^{-2})\sigma^2$. By construction, $\pmb{U}$ and $\pmb{V}$ are mutually independent conditioned on $\pmb{\theta}$.

The key idea in the second step is to use $\pmb U$ for estimation and $\pmb V$ for bandwidth selection. One can think of $\alpha$ as a measure of how much data we use for estimation \citep{brown2013poisson}. When $\alpha=0$, the entire data set is used for estimation and none for hyper-parameter calibration. This article uses $\alpha=10\%$.   
%As outlined in steps 2-5 of algorithm~\ref{algo}, the estimation procedure is conducted using $\pmb{U}$ instead of $\pmb{X}$ along with adjustments for its slightly bigger variance than $\sigma^2$. 

The steps are repeated for a list of prefixed $h$ values yielding estimates of the posterior probabilities $\{\hat p_i[h]: 1 \leq i \leq n\}$ of the latent states and marginal densities estimates $\hat f_1[h]$ and $\hat f_0[h]$. 
For each $h$ value, we estimate the MSE as 
\begin{equation}
\widehat{\text{mse}}[h]=\sum_{i=1}^{n}\left[\,\sum_{j \in \{0,1\}}\hat{q}_{i,j}[h]\left\{u_i+\sigma_u^2\,\dfrac{\hat f'_{j}[h](u_i)}{\hat f_{j}[h](u_i)}\right\}-v_i  \,\right]^2~, 
\end{equation}
where, $\hat{q}_{1,j}[h]=\hat p_1[h]$ and $\hat{q}_{0,j}[h]=1-\hat p_1[h]$. The optimal value of bandwidth is chosen as: 
$$
\hat h = \argmin_{h} \widehat{\text{mse}}[h]. 
$$
Subsequently, the HMMT estimator is computed by running the generalized BW algorithm with the bandwidth set at $\hat h$ and $\alpha=0$.

%It is to be noted that the proposed way of sample splitting offers more flexibility that the conventional leave-sample-out techniques. Due to the dependence among $x_i$s, their are not interchangeable and the conventional  \textit{leave-out} way of data splitting does not allow any natural break-point.

\subsection{Initialization and the HMMT estimator} 

Another important caveat about Algorithm~\ref{algo} is the initialization step. To prevent the algorithm getting  trapped in a local maximum,  a good initialization is important.  For this purpose we use the algorithm developed in \cite{ko2015dirichlet} for estimation in an HMM with unknown number of changing points. As the probability of the process being in control $\pi_0\geq 0.5$, $\nu^{(0)}$ is estimated by the mode of  the component with the largest probability which is subsequently attributed to $\hat f^{(0)}_0$.  The remaining states as estimated via a Dirichlet process model contributes towards the out-of-process marginal densities $\hat f^{(0)}_1$.  It was found that such an initialization produced reasonable values of $\hat{\mathcal{A}}^{(0)}$. Henceforth for simplicity, we assume $\sigma=1$. The detailed estimation procedure is summarized by Algorithm~\ref{algo} below. 

\begin{algorithm}[H]
	\caption{HMM-Tweedie Estimator}\label{algo}
	\begin{algorithmic}[1]
		\State Sample splitting. 
		
		With $\alpha=0.1$, add independent noise $\pmb Z\sim N(0,I_n)$: 
	    $\pmb U=\pmb X+\alpha\pmb Z,\ \pmb V=\pmb X-\alpha^{-1}\pmb Z$. 
	
	    Use $\pmb U$ for estimation and $\pmb V$ for bandwidth selection.

		\State  Intialization
		
		Use the algorithm in \citet{ko2015dirichlet} to get estimates $\Lambda^{(0)}=\{ \hat{\mathcal{A}}^{(0)}, \psi_0^{(0)}, \psi_1^{(0)}, \hat f_0^{(0)}, \hat f_1^{(0)}\}$.
		
		\State E step: For $t\geq 1$, compute the following in the $t$ iteration using the previous iteration estimates. 
		\begin{itemize}
			\item [  ] For $i=1,\ldots,n$ and $j, k = 0,1 $ compute
			\item [a.] The forward variable: $ \alpha_{i}^{(t)}(j)=\mathbb{P}[\Lambda^{(t-1)}](U_{1}=u_{1},...,U_{i}=u_{i}|\theta_{i}=j)$;
			\item [b.] The backward variable: $ \beta^{(t)}_{i}(j)=\mathbb{P}[\Lambda^{(t-1)}](U_{i+1}=u_{i+1},...,U_{n}=u_{n}|\theta_{i}=j) $;
			\item [c.] Posterior probabilities: \\ 
			$ \hat p_{i}^{(t)}(j)={\alpha^{(t)}_{i}(j)\beta^{(t)}_{i}(j)}/({\alpha^{(t)}_{i}(0)\beta^{(t)}_{i}(0)+\alpha^{(t)}_{i}(1)\beta^{(t)}_{i}(1)})$;\\
		    $\xi^{(t)}_{i}(j,k)=\mathbb{P}_{\Lambda^{(t-1)}}(\theta_{i}=j,\theta_{i+1}=k| \pmb U)={\hat p^{(t)}_{i}(j)\mathcal{A}^{(t-1)}_{jk} \hat{f}^{(t-1)}_{k}(u_{i+1})\beta^{(t)}_{i+1}(k)}/{ \beta^{(t)}_{i}(j)}$
		\end{itemize}
		\State M step:  Update the parameters and the non-parametric density estimate
		\begin{itemize}
			\item [a. ] $ \psi_{j}^{(t)}=n^{-1}\sum_{i=1}^n\hat p_{i}^{(t)}(j) $
			\item [b. ]
			$\mathcal{A}_{jk}^{(t)}={\sum_{i=1}^{n-1}\hat{\xi}_{i}^{(t)}(j,k)}/({\sum_{i=1}^{n-1}\hat p_{i}^{(t)}(j)}) $
			%	\item $ w_{i}^{(t)}=\dfrac{\gamma_{i}^{(t-1)}(1)}{\sum_{i=1}^{n}\gamma_{i}^{(t-1)}(1)} $
			\item [c. ] $ \hat w_{i}^{(t)}= {\hat p^{(t)}_{i}(1)}/({\sum_{i=1}^{n}\hat p^{(t)}_{i}(1)})$
		  \item [d. ] $\hat f_{1}^{(t)}[h,\pmb{u}](y)= h^{-1} \sigma_u^{-2} \sum_{i=1}^{n} {\hat w^{(t)}_{i}} K(h^{-1}\sigma_u^{-1}(y-u_{i}))$.
			\item [e. ] $\hat f_{0}^{(t)} = N(\nu^{(t)},\tau^{(t)})$
	\text{ where, }
			$(\nu^{(t)}, \tau^{(t)})=\argmax_{\nu \in \mathbb{R}, \tau \geq 0} \sum_{i=1}^n \hat p_i^{(t)}(0) \log \phi_{\tau}(u_i-\nu)$.
		\end{itemize}
		\State Iterate until the parameters and the non-parametric density estimate converges.
		\State Repeat steps 3-5 for a list of $h$ values and store the final marginal densities estimates $\hat f_0[h], \hat f_1[h]$ and the posterior probabilities $\{\hat p_i[h](j): 1 \leq i \leq n, 0 \leq j \leq 1 \}$ of the latent states.
		\State Choose bandwidth $\hat h = \argmin_{h} \widehat{\text{mse}}[h] \text{ where, }$
		
		$	\widehat{\text{mse}}[h]=\sum_{i=1}^{n}\left[\,\sum_{j \in \{0,1\}}\hat{p}_{i}[h](j)\left\{u_i+\sigma_u^2\,\dfrac{\hat f'_{j}[h](u_i)}{\hat f_{j}[h](u_i)}\right\}-v_i  \,\right]^2~.$
		\State Repeat steps 1 to 5 with bandwidth $\hat h$ and $\alpha=0$. Based on these posterior estimates $\hat{p}_i(j)$ and density estimates $\hat f_0$ and $\hat f_1$ report  location estimates:
	$
		\hat{\mu}^{\sf HMMT}_i=\sum_{j\in\{0,1\}}\hat{p}_i(j)\left(x_i+\dfrac{\hat{f}_j^{'}(x_i)}{\hat{f}_j(x_i)}\right)$.
	\end{algorithmic}
\end{algorithm}

\subsection{Operational characteristics of the new algorithm}

This section discusses the operation characteristics of Algorithm~1 that distinguishes our work from existing ones. The discussions provide intuitions that are helpful for our later theoretical analysis.

Algorithm~1 includes several modifications to the conventional Baum-Welch algorithm of \citet{baum1970maximization}. For understanding the crucial differences first consider $\alpha=1$, i.e, no sample splitting. 
The average log-likelihood for complete data $ \{\pmb x,  \pmb{\theta}  \} $ assuming $\pmb{\theta}$ known is:
\begin{equation*}
l_n(\Lambda)=n^{-1}\log\bigg(\sum_{\pmb{\theta}} P(\pmb{X},\pmb{\theta}|\Lambda)\bigg), \text{ where }  P(\pmb{X},\pmb{\theta}|\Lambda) =\psi_{\theta_{1}}\prod_{i=2}^{n}a_{\theta_{i-1},\theta_{i}}\prod_{i=1}^{n}f_{\theta_{i}}(x_{i})~.
\end{equation*}
When $ \pmb{\theta}$ is unknown, interchanging logarithm and summation above, we iteratively maximize the following lower bound $\tilde l_n(\Lambda^{(t)}| \Lambda^{(t-1)})$ of $l_n(\Lambda^t)$,
\begin{equation*}
\tilde l_n(\Lambda^{(t)}|\Lambda^{(t-1)})=Q_{n}(\Lambda^{(t)}| \Lambda^{(t-1)})+H_n(\Lambda^{(t-1)}),
\end{equation*}
where, $n H_n(\Lambda^{(t-1)})= - \mathbb{E}_{\pmb{\theta}|\pmb{X}, \Lambda^{(t-1)}}\log P(\pmb{\theta}|\pmb{X}, \Lambda^{(t-1)})$ is the entropy, and $Q_{n}$ decouples as $ Q_{n}=Q_{1,n}+Q_{2,n}$ with
$$n\,Q_{1,n}(\Lambda^{(t)}|\Lambda^{(t-1)})= \mathbb{E}_{\theta_1|\pmb{X},\Lambda^{(t-1)}}\log P(\theta_1)+\sum_{i=2}^n \mathbb{E}_{\theta_{i-1},\theta_i|\pmb{X},\Lambda^{(t-1)}} \log P(\theta_i|\theta_{i-1})$$  and $n\,Q_{2,n}(\Lambda^{(t)}|\Lambda^{(t-1)})=\sum_{i=1}^n \mathbb{E}_{\theta_i|\pmb{X},\Lambda^{(t-1)}} \log P(x_i|\theta_{i-1},f_1,f_0)$. We iteratively maximize $Q_n(\Lambda^{(t)}|\Lambda^{(t-1)})$ over $\Lambda^{(t)}$ based on previous iterate $\Lambda^{(t-1)}$. If the posterior probabilities for $[\theta_i=1]$ are updated to $\hat p_i^{(t)}$ then,  $n\,Q_{2,n}(\Lambda^t|\Lambda^{(t-1)})$ equals
\begin{align*}
\sum_{i=1}^{n}\hat p^{(t)}_{i}\log\left\{\sum_{j=1}^{n}w^t_{j}\dfrac{1}{h}K\bigg(\frac{x_{i}-x_{j}}{h}\bigg)\right\}+\sum_{i=1}^{n}(1-\hat p^{(t)}_i)\log\{\phi_{\tau}(x_i-\nu)\}.
\end{align*}
Although $Q_{2,n}$ is concave in $(w_1^t,...,w_n^t)$, it is not strongly concave. Updating $w_j^t$ by maximizing $Q_{2,n}$ over the $n-1$ dimensional hyperplane as $n \to \infty$ would not yield good solutions. 
We interchange the logarithm and summation in the first term in $Q_{2,n}$ above. The resultant $\tilde Q_{2,n}$  have a closed form maxima at  $w_j^t={\hat p^{(t)}_{j}}/{\sum_{i=1}^{n}\hat p^{(t)}_{i}}$. $f_1^{(t)}$ as demonstrated in Algorithm 1 is accordingly updated. 
Standard forward-backward procedure  \citep{rabiner1989tutorial} is  used to maximize $Q_{1,n}$ and produce updated posterior probabilities $\hat p_i^{(t)}$. For $i=1,\ldots,n$ and $j, k = 0,1 $ the forward variable: $ \alpha_{i}^{(t)}(j)=\mathbb{P}_{\Lambda^{(t-1)}}(X_{1}=x_{1},...,X_{i}=x_{i}|\theta_{i}=j)$ and 
the backward variable: $ \beta^{(t)}_{i}(j)=\mathbb{P}_{\Lambda^{(t-1)}}(X_{i+1}=x_{i+1},...,X_{n}=x_{n}|\theta_{i}=j) $ are computed. 
and the posterior probabilities are updated as:
\begin{align}\label{eq:posterior.updates}
 \hat p_{i}^{(t)}={\alpha^{(t)}_{i}(1)\beta^{(t)}_{i}(1)}/({\alpha^{(t)}_{i}(0)\beta^{(t)}_{i}(0)+\alpha^{(t)}_{i}(1)\beta^{(t)}_{i}(1)})~.
 \end{align}
The objective $Q_{1,n}+\tilde Q_{2,n}$  is bounded above and increasing at each step, therefore, our algorithm will converge. In particualr, as $n\rightarrow \infty$, the true solution is a fixed point of the algorithm. This insight is crucial for establishing the theoretical properties for the proposed HMMT estimator.

\section{Asymptotic Properties of the proposed estimator}

For establishing risk properties of our proposed shrinkage estimators, we consider the following assumption that is popularly imposed for theoretical analysis in HMM starting from the pioneering works of \citet{bickel1996inference,bickel1998asymptotic}:

\medskip
\noindent\textbf{Assumption A1.} The hidden states $\pmb{\theta}$ form a stationary ergodic Markov chain and the transition probabilities satisfy $0< a_{00}, a_{11} < 1$. The Markov chain begins in its stationary state and is reversible. \\

To understand the risk properties of the proposed estimator $\hat{\pmb{ \mu}}^{\sf HMMT}$, we consider the following two quantities: 
%For simplicity, consider $\alpha = 0$. 
\begin{align}
& f_{1,n}^{\sf or}[h](u)=\bigg({\sum_{i=1}^n \theta_i}\bigg)^{-1}\sum_{i=1}^n {\theta_i} K\bigg(\frac{u-x_i}{h}\bigg), \text{ and } \nonumber \\
& \hat f_{1,n}[h](u)=\bigg(\sum_{i=1}^n \hat p_{i,n}\bigg)^{-1}\sum_{i=1}^n \hat p_{i,n} K\bigg(\frac{u-x_i}{h}\bigg). \label{eq:form.f1}
\end{align}
which are helpful for understanding the behaviors of the proposed estimate of $f_1$.
Note that $f_{1,n}^{\sf or}[h](u)$ is an estimator that cannot be implemented in practice, since it uses knowledge of unknown $\pmb{\theta}$. By contrast, $\hat f_{1,n}[h]$ is a practical estimator that substitutes $\hat p_{i,n}$ in place of $\pmb{\theta}$. $\hat p_{i,n}$, which can be conceptualized as the estimates of posterior probabilities $P(\theta_i=1|\pmb{X})$, can be obtained as the outputs from the proposed algorithm.

Consider $\hat p_{i,n}$ defined by \eqref{eq:posterior.updates} at an arbitrary iterative step.
Barring sample splitting, the estimate of $f_1$ in every iterative step of Algorithm 1 has the functional form of $\hat f_{1,n}[h]$. We consider a class of $f_1$ that are smooth, bounded and $\gamma$-Holder continuous. 

\medskip
\noindent\textbf{Assumption A2.} $f_1$ comes from a smooth class of functions and has bounded support. $f_1$ is $\gamma$-Holder continuous, i.e.,  $f_1 \in \mathbb{H}_{\gamma}$, where 
$$\mathbb{H}_{\gamma}=\bigg\{h: \bigg|\frac{d^s}{dx^s}h(x) - \frac{d^s}{dy^s}h(y)\bigg| \leq L|x-y| \text{ for all s } \leq \gamma-1, \text{ and } x, y \in \mathbb{R},  L >0 \bigg \}.$$

It can be seen (cf. Ch. 6 of \citealp{wasserman2006all}) that the squared $L_2$ distance of the oracle kernel density estimator $f_{1,n}^{\sf or}[h]$ from the true $f_1$, $$d^2\big(f^{\sf or}_{1,n}[h],f_1\big)= \int \left\{f^{\sf or}_{1,n}[h](u)-f_1(u)\right\}^2 \, du$$
has the following asymptotic expected value:
$$\mathbb{E}_{\pmb{X}}d^2\big(f^{\sf or}_{1,n}[h],f_1\big)=O\big(h^{2\gamma}+(nh)^{-1}\big)\text{ for any h } > 0.$$
The intuition is that, under assumption A1, $\hat f_{1,n}[h]$ can be decomposed as
$$\hat f_{1,n}[h]=f_{1,n}^{\sf or}[h]+\hat R_n[h] $$
where the residual $\hat R_n[h]$ is asymptotically negligible as $n \to \infty$ provided $\hat p_{i,n}$ are asymptotically unbiased estimates of $\theta_i$. Theory underpinning this intuition is rigorously established in the proof of the next lemma. Hence 
$$
\mathbb{E}_{(\pmb{\theta},\pmb{X})} \int \hat R_n^2[h](u)\, du = O\bigg(\frac{\log n}{nh} + \frac{(\mathbb{E}_{(\pmb{\theta},\pmb{X})} \hat p_{1,n}-\psi_1)^2}{h^{2}} \bigg)\text{ as } n \to \infty,
$$
where $\psi_1=P(\theta_i=1)$ is a fixed probability whose existence is ensured the stationarity of the Markov chain 
(Assumption A1). We summarize the above discussions in the lemma below.

\begin{lemma}\label{lem:unbiasedness} 
	If $f_1$ is $\gamma$-Holder continuous and has bounded support then under Assumption A1, the integrated $L_2$ Bayes risk of non-parametric density estimators of the form \eqref{eq:posterior.updates}-\eqref{eq:form.f1} for any $h  > 0$  is
	$$\mathbb{E}_{(\theta,\pmb{X})}d^2\big(\hat f_{1,n}[h],f_1\big)=O\bigg(h^{2\gamma} + \frac{\log n}{nh} + \frac{(\mathbb{E}_{(\pmb{\theta},\pmb{X})} \hat p_{1,n}-\psi_1)^2}{h^{2}} \bigg)~ \text{ as } n \to \infty.$$
	In particular, if $\hat p_{1,n}$ are unbiased and $h_n^{\sf o}=(\log n/n)^{1/(2\gamma+1)}$ then, 
	$$\mathbb{E}_{(\theta,\pmb{X})}d^2\big(\hat f_{1,n}[h_n^{\sf o}],f_1\big)=O\{(\log n/n)^{2\gamma/(2\gamma+1)} \} \text{ as } n \to \infty.$$
\end{lemma}

The above results shows that with good posterior probability estimates, the algorithm 1 provides consistent non-parametric estimation of $f_1$. As $f_0$ is parametrized as a Gaussian density, its estimation consistency also subsequently follows. 

The free parameters in $\Lambda$ are
$\mathcal{A}=(a_{00},a_{11})$ and $\nu$, $\tau$ and $f_1$. For understanding the evolution of the estimates in Algorithm 1 from the initial solutions towards the true $\Lambda^*$, note that
under Assumption A1, the population criteria $Q_1(\Lambda|\tilde \Lambda):=\lim_{n \to \infty} Q_{1,n}(\Lambda|\tilde\Lambda)$ and $\tilde Q_2(\Lambda|\tilde \Lambda):=\lim_{n \to \infty} \tilde Q_{2,n}(\Lambda|\tilde \Lambda)$ are well-defined. Suppose our initial solution $\Lambda_n^0$ is in the vicinity of $\Lambda^*$ such that with estimates 
$\hat f^{(1)}_{0,n}$ and $\hat f^{(1)}_{1,n}$ that are improvements over the initial estimates $\hat f^{(1)}_{0,n}$ and $\hat f^{(1)}_{1,n}$ in the sense $d(\hat f^{(1)}_{1,n},f_1^*)<d(\hat f^{(0)}_{1,n},f_1^*)$ and $d(\hat f^{(1)}_{0,n},f_0^*)<d(\hat f^{(0)}_{0,n},f_0^*)$, maximizing  $Q_1(A|\hat f^{(1)}_{0,n},\hat f^{(1)}_{1,n})$ and the subsequent application of \eqref{eq:posterior.updates} produces better estimates $\hat p_{i,n}^{(1)}$ of the true posterior probabilities such that $\hat f^{(2)}_{1,n}$ given by \eqref{eq:form.f1} is better than $\hat f^{(1)}_{1,n}$ with  $d(\hat f^{(2)}_{1,n},f_1^*)<d(\hat f^{(1)}_{1,n},f_1^*)$. If this feature of the evolution is maintained over the successive iterations, the bias in the posterior probability estimates decrease in every step and $\hat f^{(1)}_{0,n}$ and $\hat f^{(1)}_{1,n}$ will ultimately converge to the true $f_0^*$ and $f_1^*$ respectively at the rate given in Lemma~\ref{lem:unbiasedness}. Having a good initial solution that has negligible asymptotic bias, which implies that successive iterations would have the aforementioned properties, is hence essential for successful convergence in Algorithm 1. Our assumption on the properties related to initialization is given next. 

\noindent\textbf{Assumption A3.} The initial solution $\mathcal{A}_0$, $f^{(0)}_0$, $f^{(1)}_1$ is in the neighborhood of the true solution. The behaviour of the population log-likelihood in this neighborhood is such that for all $t$,
 the posterior estimates $\hat{\pmb{p}}^{(t)}$ based on $\mathcal{A}^{(t)}=\argmax_\mathcal{A} Q_1(\mathcal{A}|f^{(t-1)}_0, f^{(t-1)}_1)$ produce estimates $f_1^{(t)}$ and $f_0^{(t)}$ in Algorithm 1 which satisfy:
$ \max_\mathcal{A} Q(\mathcal{A},f^{(t)}_0, f^{(t)}_1)> \max_\mathcal{A} Q(\mathcal{A},f^{(t-1)}_0, f^{(t-1)}_1).$\\

The results in Lemma~\ref{lem:unbiasedness} establish the risk optimality of our proposed $\hat{\pmb{ \mu}}^{\sf HMMT}$. To facilitate a universal optimality statement that is appropriate for both nonsparse and sparse regimes, we impose the following assumption on $\pmb{\mu}$ similar to \citet{brown2009}:

\medskip
\noindent\textbf{Assumption A4.} For any fixed positive $\delta$, we have
$n^{-\delta} \sup_{i=1,\ldots,n} |\mu_i| \to 0$ as $n \to \infty$. 

\begin{remark}
Based on our set-up in Section~\ref{sec.model} and Assumption A1, the above condition is equivalent to restricting the support of the priors $g_1$ and $g_0$ to 
$[-S_n,S_n]$ where $n^{-\delta} S_n \to 0$ as $n \to 0$ for any positive $\delta$. 
\end{remark}

For any fixed $h>0$, let $\hat{p}_i[h]$, $\hat{f}_1[h]$ and $\hat\gamma[h]$ be the final estimates of the $P(\theta_i|\pmb{X})$, $f_1$ and the mode of $f_0$ respectively, from algorithm 1. Recall that as $\sigma=1$, the estimates of the mean when the process is out of control is 
$$\hat \mu_i^{\sf OC}[h]= x_i + \dfrac{\hat{f}_{1}^{'}[h](x_i)}{\hat{f}_{1}[h](x_i)} = x_i + \dfrac{\sum_{j=1}^n \hat p_i[h] (x_j-x_i) K((x_i-x_j)/h)}{h \sum_{j=1}^n \hat p_i[h] K((x_i-x_j)/h)}~.$$
Barring sample splitting (using $\alpha =0$, not steps 7 and 8 in algorithm 1) the HMMT estimator is $\hat{\mu}_i^{\sf HMMT}[h]=(1-\hat p_i[h])\hat \nu[h]+ \hat p_i[h] \hat \mu_i^{\sf OC}[h]$. 
As the HMMT estimator involves ratio of estimates $\hat f'$ and $\hat f$, the approximation error bound in Lemma~\ref{lem:unbiasedness} on $\hat f$ does not trivally lead to optimal risk properties of  the HMMT estimator. For achieving asymptotically optimal risk performance, consider a modified robust version of the HMMT estimator that truncates very high out-of-control mean values: 
$$\hat{\mu}_i^{\sf T}[h]=(1-\hat p_i[h])\hat \nu[h]+ \hat p_i[h] \,\text{sign}(\hat \mu_i^{\sf OC}[h])\,\max(\,|\hat \mu_i^{\sf OC}[h]|,\,\sqrt{3 \log n}\,)~.$$
The following result akin to Theorem 1 of \citet{brown2009} shows that $\hat{\pmb{\mu}}^{\sf T}$ is asymptotically \textit{oracle optimal} as it can achieve the mean square error of the oracle shrinkage estimator defined in \eqref{eq.OR}. The proofs of the results in this section are provided in the supplements.

\begin{theorem}\label{thm1}
	Under Assumptions A1 to A4, for any $h_n$ such that $h_n \log n \to 0$ but $n^{\delta} h_n \to \infty$ for any positive $\delta$,  we have		
	$$\lim_{n \to \infty} \dfrac{\mathbb{E}||\pmb{\mu}_n-\hat{\pmb{\mu}}_{n}^{\sf T}||_2^2 }{\mathbb{E}||\pmb{\mu}_n-\tilde{\pmb{ \mu}}_n^{\sf OR}||_2^2}=1.$$	
\end{theorem}

\section{Simulation Studies}

This section conducts simulation studies to compare the performance of our proposed HMMT with the following four estimators:
 
\begin{description}

\item (a) {\bf GMM.3}: we use BW algorithm and Gaussian mixture models with 3 mixing components for modeling $f_1$; 
\item (b) {\bf GMM.DP}: similar to GMM.3, we use BW algorithm and Gaussian mixture models. However, the number of components in the Gaussian mixture will not be fixed as before. Instead, the number of components is estimated using the Dirichlet Process model as described in \cite{ko2015dirichlet};
\item (c) {\bf T.ND}: we ignore the Markovian dependence structure and apply Tweedie's formula. The algorithm in \citet{Fu2017nest} has been used to choose the tuning parameter.

\item (d) {\bf OR}: the oracle estimator in \eqref{eq.OR} which uses knowledge about the latent parameters. 
  
\end{description}

\subsection{Comparison of the MSEs}

In Table~\ref{tab-1} we report the mean squared errors averaged over 50 replications. We consider 11 simulation scenarios. In each scenario, we simulate $n=2,000$ observations. The latent states $\pmb{\theta}$ are generated based on a two-states Markov chain where the transition matrix has the probability of being in-control $\mathcal{A}_{00}$ fixed at $0.95$ while the probability of being out-of-control $\mathcal{A}_{11}$ is varied between $0.2$ to $0.8$. Across all the scenarios $g_0$ was fixed as the distribution with point mass at $0$. 
In cases I to VI, the out-of-control prior $g_1$ was generated from the following 6 densities:
\begin{description} 
\item (a) uniform distribution with support on $[-9,9]$;
\item (b) Asymmetric triangle distribution on $[-30,30]$ with mode at $6$;
\item (c) Levy distribution with scale $7$ on $[0,\infty]$;
\item (d) Non-central Chi-square distribution with $3$ degrees of freedom and non-centrality parameter at $2$;
\item (e) Weibull distribution with shape $2$ and scale $5$;
\item (f) Burr distribution with location $0$, scale $2$ and shape parameters $2$ and $0.5$. In cases VII to XI, we study the performance of the  estimators when $g_1$ is generated from mixture of the above densities.  
\end{description}

The following observations can be made based on the simulation results.
\begin{itemize}
\item Across all the regimes, our proposed HMMT estimator significantly improves the non-parametric shrinkage estimator T.ND which does not use dependent structure, which shows the importance of taking the dependence structure into account. 
\item In many cases, incorporating Markov structure result in efficiency gain even when the model is mis-specified.
\item The HMMT estimator also improves on the Gaussian mixture model based estimation strategies and has error rates quite close to that of the Oracle estimator in \eqref{eq.OR}. GMM.3 sometimes has higher MSE than even the non-parametric shrinkage estimator T.ND that does not use dependent structure. This reflects the usefulness of using Kernel density based non-parametric approach in HMMT.
\end{itemize}

\subsection{Comparison of the estimated $f_1$'s}

When implementing the proposed HMMT method, the estimate GMM.DP has been used in the initialization step. This indicates that, in the cases where HMMT improves significantly over GMM.DP, the proposed algorithm employs $f_1$ that is sufficiently far away from the Gaussian mixture family. This would typically happen when the distribution of out-of-control averages is heavy-tailed. 
In Figure 2, the estimated $f_1$ across these simulation regimes are displayed for the case $\mathcal{A}_{11}=0.8$. Across all the studied regimes, the HMMT estimator is evidently smoother and its shape is closer to the truth.

\begin{figure}[!h]
	\centering
	\includegraphics[scale=0.45]{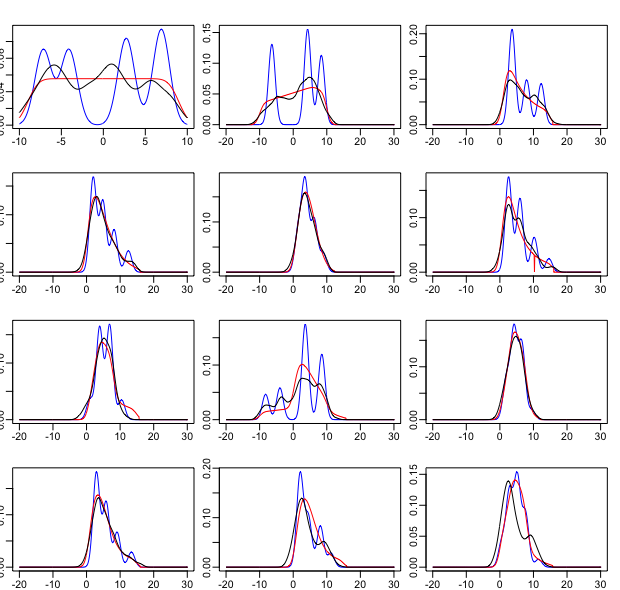}
	\caption{From top to bottom (by columns), we have the estimated density $\hat f_1$ from GMM.DP (blue) and HMMT (red) across the eleven cases considered in Table~\ref{tab-1}. The true $f_1$ is plotted in black.}
\end{figure}

\begin{table}[!ht]
	\centering
	\caption{As the probability of remaining out of control $\mathcal{A}_11$ and the out of control prior $g_1$ varies, MSE of the Tweedie estimator that does not use dependence structure (T.ND), conventional Baum-Welch algorithm using Gaussian mixture models with 3 mixing components (GMM.3) and with the number of components estimated by using Dirichlet Process model (GMM.DP) are reported along with the performance of our proposed HMMT estimator and that of the Oracle estimator in \eqref{eq.OR}.}\label{tab-1}
	\begin{tabular}{|c|c|c|c|c|c|c|c|}
		\multicolumn{1}{l}{\;\;Scenarios}  & 	\multicolumn{1}{l}{\;\;Out-of-control prior $g_1$}  & \multicolumn{1}{l}{$\mathcal{A}_{11}$} & \multicolumn{1}{l}{T.ND} & \multicolumn{1}{l}{GMM.3} &\multicolumn{1}{l}{GMM.DP\,} & \multicolumn{1}{l}{HMMT\;\;} & \multicolumn{1}{l}{Oracle} \\
		\hline
	\multirow{4}[0]{*}{I} &
	\multirow{4}[0]{*}{Unif[-9,9]} 
		   & 0.2   & 0.345 & 0.213 &0.137 & 0.130   & 0.112 \\

&	  & 0.4   & 0.361  & 0.217 &0.150& 0.138 & 0.131 \\
&		  & 0.6   & 0.439 & 0.317 &0.190& 0.177 & 0.168 \\
&		   & 0.8   & 0.429  & 0.502 &0.263& 0.253  & 0.238 \\
		\hline
			\multirow{4}[0]{*}{II} &
		\multirow{4}[0]{*}{Triangle} 
		  & 0.2   & 0.378 & 0.224 &0.134 &0.124 & 0.110 \\
	&	    & 0.4   & 0.294 & 0.242 &0.154 &0.136 & 0.122 \\
	&	  & 0.6   & 0.319 & 0.336 &0.190 &0.176 & 0.164 \\
	&	  & 0.8   & 0.515 & 0.579 &0.284 &0.258 & 0.250 \\
		\hline
		\multirow{4}[0]{*}{III} &
		\multirow{4}[0]{*}{Levy}
		  & 0.2   & 0.313  & 0.175 &0.128& 0.120 & 0.116 \\
	&	  & 0.4   & 0.309 & 0.192 &0.144 &0.135 & 0.132 \\
	&	  & 0.6   & 0.356 & 0.215 &0.171 &0.163 & 0.160 \\
	&	  & 0.8   & 0.468 & 0.375  &0.240& 0.235 & 0.232 \\
		\hline
		\multirow{4}[0]{*}{IV} &
		\multirow{4}[0]{*}{Non-central $\chi^2$} 
		   & 0.2   & 0.358 & 0.195 &0.123& 0.117 & 0.110 \\
		 &    & 0.4   & 0.330 & 0.199 &0.151 &0.149 & 0.141 \\
		 &  & 0.6   & 0.340 & 0.246 & 0.171&0.167 & 0.159 \\
		  &  & 0.8   & 0.521 & 0.392 & 0.232&0.225 & 0.223 \\
		\hline
		\multirow{4}[0]{*}{V} &
		\multirow{4}[0]{*}{Weibull} 
		   & 0.2   & 0.352 & 0.174 &0.134& 0.130 & 0.124 \\
	&	     & 0.4   & 0.370 & 0.184 &0.149 &0.143 & 0.136 \\
	&	   & 0.6   & 0.402 & 0.195 &0.165 &0.162 & 0.160 \\
	&	  & 0.8   & 0.447 & 0.289 &0.239 &0.233 & 0.229 \\
		\hline
		\multirow{4}[0]{*}{VI} &
		\multirow{4}[0]{*}{Burr} 
		  & 0.2   & 0.329 & 0.191 &0.128 &0.122 & 0.116 \\
&	   & 0.4   & 0.365 & 0.200 &0.152 &0.150 & 0.142 \\
&		   & 0.6   & 0.428 & 0.247 &0.176 &0.174 & 0.167 \\
&	  & 0.8   & 0.421 & 0.377 & 0.233&0.228  & 0.217 \\
		\hline
		\multirow{4}[0]{*}{VII} &
				\multirow{4}[0]{*}{0.4 Unif[3,8] + 0.6 Levy} 
				  & 0.2   & 0.363 & 0.180&0.121 & 0.118   & 0.111 \\
&				   & 0.4   & 0.394  & 0.173 &0.128& 0.127 & 0.121 \\
&			  & 0.6   & 0.389 & 0.248 &0.169& 0.166 & 0.165 \\
&			  & 0.8   & 0.506  & 0.367 &0.230& 0.223  & 0.221 \\
				\hline
					\multirow{4}[0]{*}{VIII} &
				\multirow{4}[0]{*}{0.4 Non-central $\chi^2$ + 0.6 Triangle} 
			  & 0.2   & 0.340 & 0.253&0.109& 0.106 & 0.102 \\
&			    & 0.4   & 0.428 & 0.268 &0.139& 0.132 & 0.124 \\
&			  & 0.6   & 0.435 & 0.333 &0.176& 0.166 & 0.160 \\
&			   & 0.8   & 0.541 & 0.591&0.239 & 0.228 & 0.226 \\
				\hline
					\multirow{4}[0]{*}{IX} &
				\multirow{4}[0]{*}{0.4 Unif[3,8] + 0.6 Weibull}
				  & 0.2   & 0.262  & 0.153 &0.132& 0.123 & 0.112 \\
	&		   & 0.4   & 0.345 & 0.163 &0.136& 0.135 & 0.126 \\
	&		  & 0.6   & 0.395 & 0.175 &0.156& 0.153 & 0.146 \\
	&			  & 0.8   & 0.437 & 0.257  &0.219& 0.217 & 0.210 \\
				\hline
					\multirow{4}[0]{*}{X} &
				\multirow{4}[0]{*}{0.5 Weibull + 0.5 Levy} 
				   & 0.2   & 0.301 & 0.177 &0.123& 0.119 & 0.106 \\
	&			     & 0.4   & 0.321 & 0.196 &0.142& 0.138 & 0.129 \\
	&		 & 0.6   & 0.362 & 0.226 &0.170& 0.165 & 0.156 \\
	&			   & 0.8   & 0.462 & 0.375 &0.240& 0.238 & 0.231 \\
				\hline
					\multirow{4}[0]{*}{XI} &
				\multirow{4}[0]{*}{0.5 Non-central $\chi^2$ + 0.5 Burr} 
				   & 0.2   & 0.295 & 0.176&0.122 & 0.120 & 0.112 \\
		&	    & 0.4   & 0.329 & 0.205 &0.150 &0.147 & 0.138 \\
		&		  & 0.6   & 0.370 & 0.276&0.178 &0.174 & 0.166 \\
		&		 & 0.8   & 0.456 & 0.404 &0.244 &0.242 & 0.235 \\
				\hline
								\multirow{4}[0]{*}{XII} &
				\multirow{4}[0]{*}{0.6 Non-central $\chi^2$ + 0.4Unif[3,8] } 
				& 0.2   & 0.332 & 0.136&0.123 & 0.115 & 0.107 \\
				&	    & 0.4   & 0.359 & 0.160 &0.150 &0.139 & 0.132 \\
				&		  & 0.6   & 0.407 & 0.209&0.185 &0.185 & 0.180 \\
				&		 & 0.8   & 0.473 & 0.307 &0.236 &0.205 & 0.202 \\
				\hline
	\end{tabular}%
	\label{tab:addlabel}%
\end{table}%

\clearpage

\section{Real data analysis}

This section we illustrate our methodology by applying it to several real data analyses. 

\subsection{Copy number variation}

\begin{figure}[H]
	
	\includegraphics[scale=0.4]{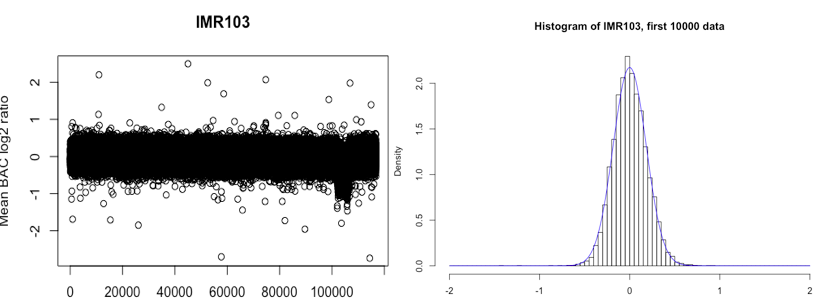}
	\caption{Left panel: The IMR103 data, the x-axis is the gene number, y-axis is the mean BAC $\log 2$ ratio.Right panel: Histogram of the first 10000 genes' mean BAC $\log 2$ ratio.}\label{cnv1}
\end{figure}

We analyze the IMR103 data in \cite{sharp2006discovery}, which are displayed in Figure \ref{cnv1}. The gene copy number is the number of copies of a particular gene in the genotype of an individual. It is widely believed that in healthy cells, the average copy number should be 2. A shift away from 2 is a genomic disorder and is usually related to certain disease. It is clear from the left panel of Figure \ref{cnv1} that in the region from 100000 to 110000, there is a shift away from 0 in $ log_{2}$ ratio. We take $ g_{0}=\delta_{0}$ and model the data using an HMM. The two hidden states 0 and 1 can be interpreted as healthy/unhealthy genes. The noise variance is estimated as the sample variance of the the first 10000 data (0.183). The right panel of Figure \ref{cnv1} shows the histogram of the first 10000 data along with the density function of $N(0,0.183^{2})$. We will take this as $ f_{0}$. 

For the data analysis, we focus on the data from position position 111000 to 112999 (Call it $\pmb{U}$) and  from position 1113000 to 117999 (Call it $\pmb{V}$). For each data in $\pmb{U}$ we add a random noise $\epsilon_1\sim N(0.0.183^2)$. Define $\pmb{U}_1=\pmb{U}+\pmb{\epsilon_1}$, For each data in $\pmb{V}$ we also add a noise from the same distribution $\epsilon_2\sim N(0,0.183^2)$. Let $\pmb{V}_1=\pmb{V}+\pmb{\epsilon_2}$ and $\pmb{V}_2=\pmb{V}-\pmb{\epsilon_2}$. Under this construction, $\pmb{V}_1$ and $\pmb{V}_2$ are independent with the same mean. We will construct the estimators based on $\pmb{U}_1$, and estimate the mean vector of $\pmb{V}_1$ (call it $\hat{\pmb{\mu}}$). And use the average of $\|\hat{\pmb{\mu}}-\pmb{V}_2\|_2 ^2$ as prediction error.
The plot of $\pmb{U}$ and $\pmb{V}$ are shown in figure 14 and figure 15 respectively.
%We wil as $ \pmb{x_{1}}^{T}$. This part of the data is shown in figure 10. Its histogram is shown in figure 11. We will use the segment from 101500 to 102500 as $ \pmb{x_{1}}^{CV} $, shown in figure 12 and 13. 
\begin{figure}[H]
	
	\includegraphics[scale=0.4]{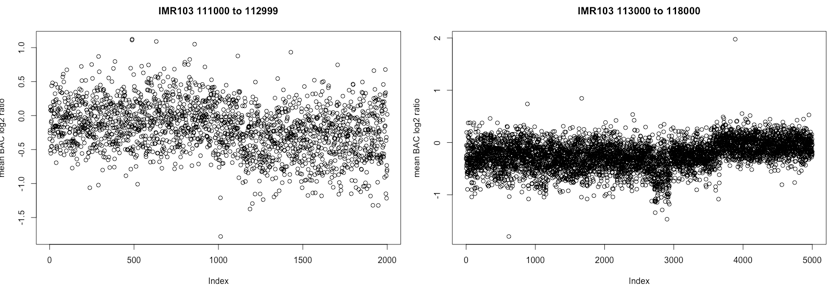}
	\caption{Left panel: The IMR103 data gene number 111000 to 112999, the x-axis is the gene number, y-axis is the mean BAC $\log 2$ ratio.Right panel: The IMR103 data gene number 113000 to 118000, the x-axis is the gene number, y-axis is the mean BAC $\log 2$ ratio.}\label{cnv2}
\end{figure}

\begin{figure}[H]
	
	\includegraphics[scale=0.4]{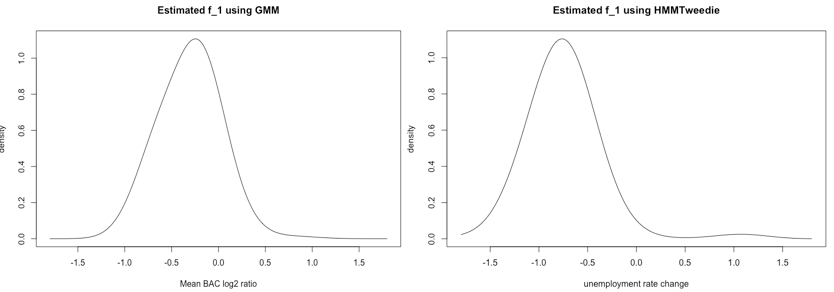}
	\caption{Left panel: Estimated $f_1$ for mean BAC $\log 2$ ratio using GMM. Right panel: Estimated $f_1$ for mean BAC $\log 2$ ratio using HMMTweedie.}\label{cnv3}
\end{figure}

The estimated $f_1$ using the Gaussian mixture model and our non-parametric method { are displayed in Figure \ref{cnv3}}. Use the sample splitting method we discussed in Section 3.2, the bandwidth is chosen to be 0.28. Finally we apply the GMM.DP and HMMT methods to the data set. The prediction error for GMM.DP is 0.130, whereas the prediction error of the proposed HMMT is 0.121. This illustrates the benefit of using the nonparametric approach to modeling $f_1$.

\subsection{Internet search trend}

\begin{figure}[H]
	
	\includegraphics[scale=0.4]{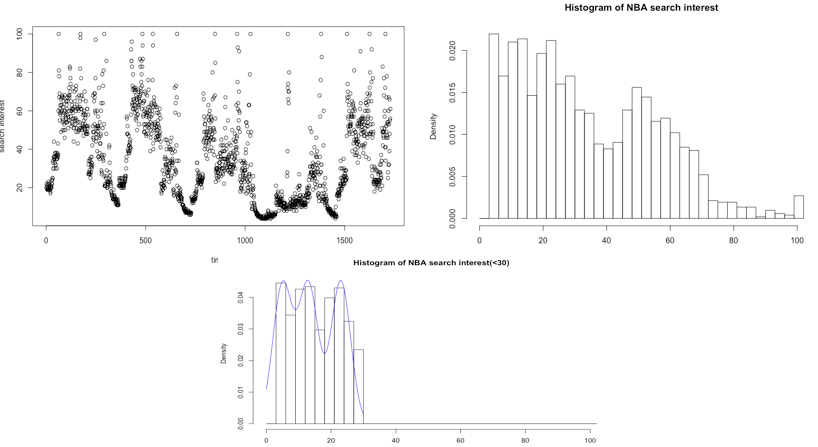}
	\caption{Top left panel: Plot for the search interest score, x-axis is time, y-axis is the google search interest score for the keyword "NBA". Top right panel: Histogram of the search interest score for the keyword  "NBA". Bottom panel: Histogram of the search interest score (<30) for the keyword  "NBA"}\label{nba1}
\end{figure}

\begin{figure}[H]
	
	\includegraphics[scale=0.4]{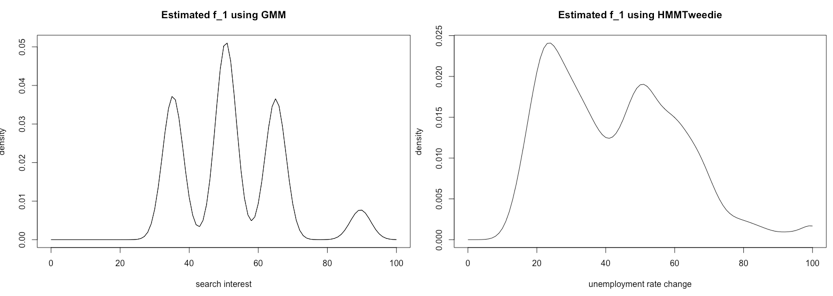}
	\caption{Left panel: Estimated $f_1$ using GMM. Right panel: Estimated $f_1$ using HMMTweedie }\label{nba2}
\end{figure}

This section applies the proposed methods for analysis of search trend data. The key word  of interest is ``NBA''. The data are collected from \texttt{Google Trend} for the period 08/29/2013 to 05/28/2018. According to Google, ``Numbers represent search interest relative to the highest point on the chart for the given region and time. Hence a value of 100 is the peak popularity for the term. A value of 50 means that the term is half as popular. A score of 0 means there was not enough data for this term.'' Because of the increasing accessibility of the Internet, if we directly take the time interval to be from 08/29/2013 to 05/28/2018, there will be a clear increasing trend. To adjust for that, we collect the data by taking the time window to be 4 months. Since the numbers are relative, homoscedastic errors seem to be a reasonable assumption.  Figure \ref{nba1} display the search trend data and its histogram. Bottom panel of  Figure \ref{nba1} shows the histogram of the data with value less than 30, where the line represents the density function of $ \frac{1}{3}N(5,3)+ \frac{1}{3}N(13,3)+\frac{1}{3}N(25,3)$
The two hidden states 0 and 1 can be interpreted as no event/event. An event could be an important game (all-stars,finals etc.). To estimate the homoscedastic error, we compute the standard deviation of the data from 08/29/2013 to 10/02/2013, since it is during the off-season, the search trend should be stable. It turns out the standard deviation is around 3. We are primarily interested in the time when search interest is high. Thus we use $f_0$ to model those time when search interest is low. By inspection, We take $ f_0 $ to be the density for $ \frac{1}{3}N(5,3)+ \frac{1}{3}N(13,3)+\frac{1}{3}N(25,3)$. The estimated $ f_1 $ is displayed in Figure \ref{nba2}. For the proposed HMMT method, the bandwidth is chosen to be 3.
Using the estimated transition matrix and $f_1$, we can calculate the expected proportion of days when the search interest is greater than 80. The Gaussian mixture model gives an estimate of 0.0294, the non-parametric model gives an estimate of 0.0266, whereas the proportion obtained from the data is 0.0243. We can see that the HMMT method provides a better estimate.

\subsection{Change in unemployment rate}

\begin{figure}[H]
	
	\includegraphics[scale=0.4]{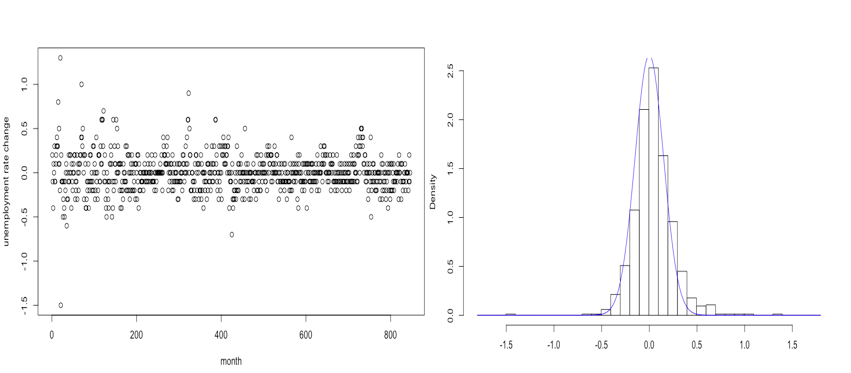}
	\caption{Left panel: Plot of unemployment rate change from February 1948 to August 2018. x-axis is time, y-axis is unemployment rate change. Right panel: Histogram of of unemployment rate change from February 1948 to August 2018}\label{unem1}
\end{figure}

Finally we illustrate the method using the unemployment data. Left panel of Figure \ref{unem1} is a plot of monthly unemployment rate change from February 1948 to August 2018. The data is from U.S. Bureau of Labor Statistics website. Since we are only considering the \emph{changes} of unemployment rate, it is reasonable to assume that the errors are homoscedastic.
We can see from the left panel of Figure \ref{unem1} that most of the time the change in unemployment rate is close to 0, and dramatic change often appears in clusters, exhibiting dependence structure that can be reasonably described by an HMM. 

\begin{figure}[H]
	
	\includegraphics[scale=0.4]{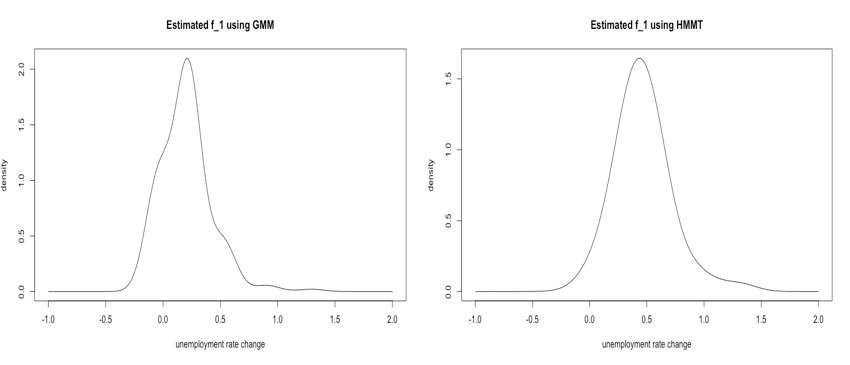}
	\caption{Left panel: Estimated $f_1$ for unemployment rate change using GMM. Right panel: Estimated $f_1$ for unemployment rate change using HMMTweedie}\label{unem2}
\end{figure}

Suppose we are interested in the months where there is a big increase in unemployment rate. Economists and policy makers may want to focus on those times and try to figure out the possible causes to avoid increase in unemployment rate in the future. The right panel of Figure \ref{unem1} shows a histogram of the unemployment rate change when it is $\leq 0.2\%$, together with the density function for $0.8N(0,0.11)+0.2N(-0.25,0.11)$. We can see the density function matches the data quite well, therefore, we assume $f_0$ to be $0.8N(0,0.11)+0.2N(-0.25,0.11)$
The estimated $ f_1 $ is given in Figure \ref{unem2}. For the proposed HMMT, we choose the bandwidth to be 0.15. The Gaussian mixture model assumes there are 6 components. However, as we can see from the histogram, the mixture model is inadequate. By contrast, the proposed nonparametric model describes the data quite well; this clearly shows the flexibility and benefit of the HMMT method.

\begin{figure}[!h]
	\includegraphics[scale=0.4]{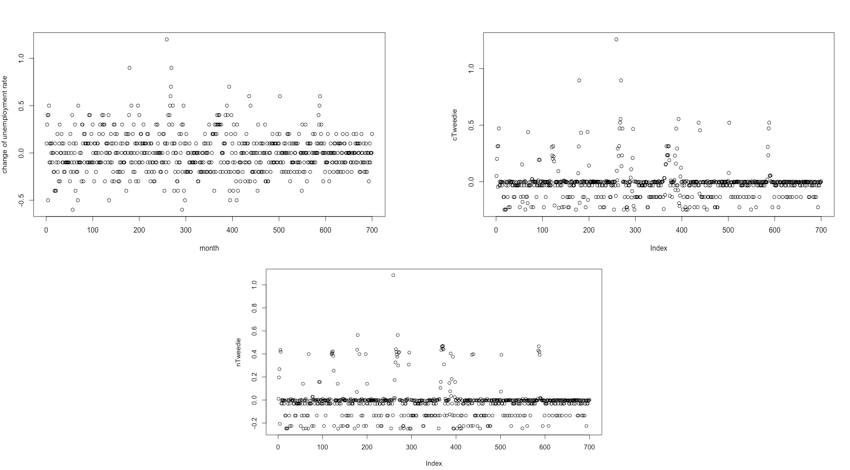}
	\caption{Top left panel: Plot of change of unemployment rate January 1960 to June 2018. Top right panel: Estimated true change of unemployment rate January 1960 to June 2018 using GMM. Bottom: Estimated true change of unemployment rate January 1960 to June 2018 using HMMTweedie.}\label{unem3}
\end{figure}

Next we apply our GMM.DP and HMMT to analyze the unemployment rate data (monthly, seasonal adjusted, January 1960 to June 2018), which are displayed in Figure \ref{unem3}. The data are obtained from the website of Federal Reserve Bank of St. Louis. Top right panel of Figure \ref{unem3} presents the estimated true change of unemployment rate using a Gaussian mixture model. Bottom panel of Figure \ref{unem3} is the estimate obtained using the proposed HMMT. We can see that HMMT has a clear ``de-noise'' effect on the data. The HMMT estimate is less scattered and more robust than GMM.DP. This can help researchers and policy makers to have a better idea of the true change of unemployment rate, and focus on the studies on the policies over the months that have a real impact on the unemployment rate.

\section{Discussion}
We developed a non-parametric estimator for estimating mean vector under Markovian dependence. 
It will be interesting to introspect if following the bivariate kernel density approach in \citet{Fu2017nest} the proposed methodology can be extended to heteroskedastic Markov set-ups with known variances.  
Developing shrinkage algorithms  when the variance  in the observations is unknown and needs to jointly estimated from the data will be important. Estimation in multivariate HMMs where the observed outputs $X_i$'s are vectors finds wide applications \citep{kale2003towards,fiecas2017shrinkage}. In this context it will be useful to study shrinkage estimation particularly in the presence of latent structual properties on the covariances. 

\section{Acknowledgements}
G. Mukherjee's work was partly supported by the NSF grant DMS-1811866. W. Sun's work
was partly supported by the NSF grant DMS-1712983.

\bibliographystyle{spbasic}

\bibliography{hmmref,mybib,pred-inf,pred-inf-1}
\clearpage

\printauthorindex

\printindex

\clearpage
\newpage
\setcounter{page}{1}
\section*{Supplement}
Here,  we provide the proofs of the results in Section~4.
\subsection*{Proof of Lemma \ref{lem:unbiasedness}}
Note that it is sufficient to bound $\mathbb{E}_{(\theta,\pmb{X})}d^2\big(\hat f_{1,n}[h],f_{1,n}^{\sf or}[h]\big).$ The result then follows from the triangle inequality. Let $\mathcal{B}$ be the event that $|\sum_{i=1}^{n}\theta_i-\psi_{1}n|<\frac{1}{2}\psi_{1}n$, throughout the proof we assume $\mathcal{B}$ holds, as by Hoeffding's inequality, $\mathbb{P}(\mathcal{B}^c)=O(e^{-n/2})$.
Write $b_j=\theta_{j}/\sum_{i=1}^{n}\theta_{i}$, $b^*_j=\hat{p}_{j,n}/\sum_{i=1}^{n}\hat{p}_{i,n}$, and $\phi_h(x-x_j)=\frac{1}{h}K(\frac{x-x_j}{h})$ then
\begin{align*}
\left\{	\hat f_{1,n}[h](x)-f_{1,n}^{\sf or}[h](x)\right\}^2&=\sum_{j=1}^{n}(b^*_j-b_j)^2\phi^2_h(x-x_j)\\
&+\sum_{j\neq k}^{}(b^*_j-b_j)(b^*_k-b_k)\phi_h(x-x_j)\phi_h(x-x_k).
\end{align*}
\newcommand{\Var}{\textsf{Var}}
We first bound $\mathbb{E}(b^*_j-b_j)^2$. Write $\mathbb{E}(b^*_j-b_j)^2=\{\mathbb{E}(b^*_j-b_j)\}^2+\Var(b^*_j-b_j)$. 
It is clear that $\mathbb{E}(b^*_j)$ and $\mathbb{E}(b_j)$ are both of order $O(n^{-1})$. Hence $\{\mathbb{E}(b^*_j-b_j)\}^2=O(n^{-2})$. Next consider $\Var(b^*_j-b_j)=\Var(b^*_j)+\Var(b_j)-2Cov(b^*_j,b_j)$. We have
$$
\Var(b^*_j)=\Var\bigg\{\dfrac{\theta_j}{\sum_{i=1}^{n}\theta_{i}}\bigg\}	\leq  \mathbb{E}\bigg\{\dfrac{1}{\sum_{i=1}^{n}\theta_{i}}\bigg\}^2=O(n^{-2}).
$$
Similarly $\Var(b_j)=O(n^{-2}).$ It follows from Cauchy-Schwarz inequality that $Cov(b^*_j,b_j)=O(n^{-2}).$ Therefore $\Var(b^*_j-b_j)=O(n^{-2})$ and $\mathbb{E}(b^*_j-b_j)^2=O(n^{-2})$. Using the fact that $\int\phi_{h}^2(x-x_j)dx=O(h^{-1})$, we have
\begin{equation}\label{step21}
\int\mathbb{E} \sum_{j=1}^{n}(b^*_j-b_j)^2\phi_{h}^2(x-x_j)dx=O\{(nh)^{-1}\}.
\end{equation}

Next we bound $\mathbb{E}\{(b^*_j-b_j)(b^*_k-b_k)\}$ for $j\neq k$. Consider the decomposition 
\begin{equation}\label{decomp2}
\mathbb{E}\{(b^*_j-b_j)(b^*_k-b_k)\}=\mathbb{E}(b^*_j-b_j)\mathbb{E}(b^*_k-b_k)+Cov(b^*_j-b_j,b^*_k-b_k).
\end{equation}
%We first show that $\mathbb{E}\{(b^*_j-b_j)(b^*_k-b_k)\}=O(n^{-3}\log n)$. 
Note that 
$$
\mathbb{E}b_j=\psi_{1}\mathbb{E}\left(\dfrac{1}{\sum_{i=1}^{n}\theta_{i}}\right)+Cov\left(\theta_j,\dfrac{1}{\sum_{i=1}^{n}\theta_{i}}\right).
$$
Assumption A1 implies $Cov(\theta_j,\theta_k)=O(\gamma^{|j-k|})$ for some $\gamma>0$. Hence for every $j$ we can focus on its $\log n$ neighborhood, it follows that $$\mathbb{E}\left(1/\sum_{i=1}^{n}\theta_{i}|\theta_{j}=1\right)-\mathbb{E}\left(1/\sum_{i=1}^{n}\theta_{i}|\theta_{j}=0\right)=O(n^{-2}\log n).$$
Some elementary calculation shows $Cov\left(\theta_j,\dfrac{1}{\sum_{i=1}^{n}\theta_{i}}\right)=O(n^{-2}\log n).$
By Lemma 3 from \cite{fu2019information}, together with the Markov structure, we have $$\mathbb{E}\left(\dfrac{1}{\sum_{i=1}^{n}\theta_{i}}\right)=\dfrac{1}{\psi_{1}n}+O(n^{-2}\log n ).$$
Hence $\mathbb{E}b_j=1/n+O(n^{-2}\log n ).$
Note that $\mathbb{E}b^*_j=1/n+O\left(\frac{\mathbb{E} \hat p_{1,n}-\psi_1}{n}+\frac{\log n}{n^2}\right).$ It follows that $\mathbb{E}(b^*_j-b_j)\mathbb{E}(b^*_k-b_k)=O\left(\frac{\mathbb{E} (\hat p_{1,n}-\psi_1)^2}{n^2}+\frac{\log^2 n}{n^4}\right)$. By Lemma 4 from \cite{fu2019information} and the Markov structure we also have $Cov(b^*_j-b_j,b^*_k-b_k)=O(n^{-3}\log n)$. Hence,
$$\int \mathbb{E} \sum_{j\neq k}^{}(b^*_j-b_j)(b^*_k-b_k)\phi_h(x-x_j)\phi_h(x-x_k) dx= O\bigg(\frac{\log n}{nh} + \frac{(\mathbb{E}_{(\pmb{\theta},\pmb{X})} \hat p_{1,n}-\psi_1)^2}{h^{2}} \bigg). $$
The lemma follows.

\subsection*{Proof of Theorem \ref{thm1}}
First, notice that as a consequence of A1, 
$\mathbb{E}||\pmb{\mu}-\hat{\pmb{\mu}}^{OR}(\pmb x)||_2^2=O(n). $
Thus the conditions of theorem 1 in \citealp{brown2009} are satisfied. Our theorem is an adaption of theorem 1 in \citealp{brown2009}. The proof follows the same logic, we provide a sketch.\\
We will show $\mathbb{E}||\hat{\pmb{\mu}}^T_n-\tilde{\pmb{ \mu}}_n^{\sf OR}||_2^2=o(n^\epsilon) $ for any $\epsilon$>0.
Let $p_i=\mathbb{P}(\theta_i=1|\pmb x)$. Note that it follows from the proof of Lemma \ref{lem:unbiasedness} that $\hat{p}_i-p_i=o(1)$. This fact together with Lemma 2 in \cite{brown2009} implies we only need to show
\begin{equation}\label{eq1}
\mathbb{E}\sum_{i=1}^{n}\left\{ \dfrac{f'_1[h](X_i)}{f_1[h](X_i)}-\dfrac{\hat{f}'_1[h](X_i)}{\hat{f}_1[h](X_i)}    \right\}^2 =o(n).
\end{equation}
Where $f_{1}[h](x)=\int \dfrac{1}{\sqrt{1+h}}\phi\bigg(\dfrac{x-\mu}{\sqrt{1+h}} \bigg)dg_1(\mu). $
Follow the proof of lemma 3 in \citealp{brown2009} we first show for an independent sample $\tilde{\pmb{X}}=(\tilde{X}_1,...,\tilde{X}_n), \ \tilde{X}_i\sim N(\mu_i,1),\ i=1,...n$ we have 
\begin{equation}\label{indepsample}
\mathbb{E}\sum_{i=1}^{n}\left\{ \dfrac{f'_1[h](\tilde{X}_i)}{f_1[h](\tilde{X}_i)}-\dfrac{\hat{f}'_1[h](\tilde{X}_i)}{\hat{f}_1[h](\tilde{X}_i)} \right\}^2  =o(n).
\end{equation}
We can write $\frac{\hat{f}'_1[h](x_i)}{\hat{f}_1[h](x_i)} =\frac{f'_1[h](x_i)+R_1}{f_1[h](x_i)+R_2}$. Then on the region $\mathcal{R}$ defined in the proof of lemma 3 in in \citealp{brown2009},
$$\mathbb{E}\sum_{i=1}^{n}\left\{ \dfrac{f'_1[h](\tilde{X}_i)}{f_1[h](\tilde{X}_i)}-\dfrac{\hat{f}'_1[h](\tilde{X}_i)}{\hat{f}_1[h](\tilde{X}_i)} \right\}^2=O\left\{\mathbb{E}\bigg(  \dfrac{R_1}{f_{1}[h](\tilde{X}_i)+R_2}\bigg)^2+  \mathbb{E}\bigg(  \dfrac{CR_2}{f_{1}[h](\tilde{X}_i)+R_2}\bigg)^2       \right\}.$$
A consequence of Lemma \ref{lem:unbiasedness} is that $\mathbb{E}(R_i)\rightarrow 0,\ i=1,2$ as $n\rightarrow \infty$. We only need to bound The variance of $R_i$. The variances of $R_1,\ R_2$ are the variances of the density estimators $\hat{f}'_{1}[h](\tilde{X}_i)$ and $\hat{f}_{1}[h](\tilde{X}_i)$. Notice that the variances of $f_{1,n}^{\sf or}[h](\tilde{X}_i)$ and $f_{1,n}^{\sf or}[h](\tilde{X}_i)$ satisfies (55) in \citealp{brown2009}. Since $R_1\leq {\Var}f_{1,n}^{\sf or}[h](\tilde{X}_i)$ and $R_2\leq {\Var}f_{1,n}^{\sf or}[h](\tilde{X}_i)$, $R_1$, $R_2$ also satisfy (55). Using Bernstein, 
$$\mathbb{P}\{R_2-\mathbb{E}(R_2)<-0.5f_{1}[h](\tilde{X}_i)\}<1/4C^2 ,$$ Since $\mathbb{E}(R_2)\rightarrow 0$, then as $n\rightarrow\infty$,
$$\mathbb{P}\{R_2<-0.5f_{1}[h](\tilde{X}_i)\}\leq1/4C^2 . $$
Use the same argument as in the proof of Lemma 3 in in \citealp{brown2009}, (\ref{indepsample}) follows. To show (\ref{eq1}), we use the same  argument in \citealp{brown2009} . Since the Markov structure at most contribute a factor of $\log n$, the conclusion is not affected. (\ref{eq1}) and Lemma 2 in \citealp{brown2009}  together with Cauchy-Schwarz inequality proves the theorem.

\end{document}